\documentclass[conference]{IEEEtran}
\IEEEoverridecommandlockouts

\usepackage{mathptmx}
\usepackage{cite}
\usepackage{amsmath,amssymb,amsfonts}
\usepackage{algorithmic}
\usepackage{graphicx}
\usepackage{textcomp}
\usepackage{xcolor}
\usepackage{comment}
\usepackage{multirow}
\usepackage[colorlinks,urlcolor=blue,citecolor=blue]{hyperref}
\usepackage{doi}
\usepackage{xurl}

\begin{document}

\title{QCaMP:  A 4-Week Summer Camp Introducing High School Students to Quantum Information Science and Technology\\

\thanks{QCaMP is funded by the U.S. Department of Energy, Office of Science, Workforce Development for Teachers and Scientists, Pathway Summer Schools. Additional support is acknowledged from Sandia National Laboratories, Lawrence Berkeley National Lab, Los Alamos National Lab, and the National Energy Research Scientific Computing Center. SAND2025-04894C}
}

\author{\IEEEauthorblockN{Megan Ivory}
\IEEEauthorblockA{\textit{Photonic Microsystems Technology} \\
\textit{Sandia National Laboratories}\\
Albuquerque, NM, USA \\
mkivory@sandia.gov}
\and
\IEEEauthorblockN{Jan Balewski}
\IEEEauthorblockA{\textit{National Energy Research}\\
\textit{Scientific Computing Center} \\
\textit{Lawrence Berkeley National Laboratory}\\
Berkeley, CA, USA \\
balewski@lbl.gov}
\and
\IEEEauthorblockN{Alisa Bettale}
\IEEEauthorblockA{\textit{Office of Government and} \\
\textit{Community Relations}\\
\textit{Lawrence Berkeley National Laboratory}\\
Berkeley, CA, USA \\
abettale@lbl.gov}
\and
\IEEEauthorblockN{Jaden Brewer}
\IEEEauthorblockA{\textit{Quantum Photonics Lab} \\
\textit{Northern Arizona University}\\
Flagstaff, AZ, USA \\
mjb739@nau.edu}
\and
\IEEEauthorblockN{Rachel Boren}
\IEEEauthorblockA{\textit{College of Health, Education,}\\
\textit{and Social Transformation} \\
\textit{New Mexico State University}\\
Las Cruces, NM, USA \\
rboren@nmsu.edu}
\and
\IEEEauthorblockN{Daan Camps}
\IEEEauthorblockA{\textit{National Energy Research}\\
\textit{Scientific Computing Center} \\
\textit{Lawrence Berkeley National Laboratory}\\
Berkeley, CA, USA \\
dcamps@lbl.gov}
\and
\IEEEauthorblockN{Lisa Hackett}
\IEEEauthorblockA{\textit{Photonic Microsystems Technology} \\
\textit{Sandia National Laboratories}\\
Albuquerque, NM, USA \\
lahacke@sandia.gov}
\and
\IEEEauthorblockN{Martin Juarez}
\IEEEauthorblockA{\textit{Bourns College of Engineering} \\
\textit{University of California Riverside}\\
Riverside, CA, USA \\
mjuar044@ucr.edu}
\and
\IEEEauthorblockN{Alina Kononov}
\IEEEauthorblockA{\textit{Quantum Algorithms and Applications} \\
\textit{Collaboratory} \\
\textit{Sandia National Laboratories}\\
Albuquerque, NM, USA \\
akonono@sandia.gov}
\and
\IEEEauthorblockN{Kan-Heng Lee}
\IEEEauthorblockA{\textit{Applied Mathematics and}\\
\textit{Computational Research Division} \\
\textit{Lawrence Berkeley National Laboratory}\\
Berkeley, CA, USA \\
KH.Lee@lbl.gov}
\and
\IEEEauthorblockN{Maryanne Long}
\IEEEauthorblockA{\textit{College of Health, Education,}\\
\textit{and Social Transformation} \\
\textit{New Mexico State University}\\
Las Cruces, NM, USA \\
long2@nmsu.edu}
\and
\IEEEauthorblockN{Andy Mounce}
\IEEEauthorblockA{\textit{Center for Integrated Nanotechnology} \\
\textit{Sandia National Laboratories}\\
Albuquerque, NM, USA \\
amounce@sandia.gov}
\and
\IEEEauthorblockN{Susan Mniszewski}
\IEEEauthorblockA{\textit{Information Sciences} \\
\textit{Los Alamos National Laboratory}\\
Los Alamos, NM, USA \\
smm@lanl.gov}
\and
\IEEEauthorblockN{Ravi K. Naik}
\IEEEauthorblockA{\textit{Applied Mathematics and}\\
\textit{Computational Research Division} \\
\textit{Lawrence Berkeley National Laboratory}\\
Berkeley, CA, USA \\
rnaik24@lbl.gov}
\and
\IEEEauthorblockN{Scott Pakin}
\IEEEauthorblockA{\textit{Applied Computer Science} \\
\textit{Los Alamos National Laboratory}\\
Los Alamos, NM, USA \\
pakin@lanl.gov}
\and
\IEEEauthorblockN{Ernesto Sanchez}
\IEEEauthorblockA{\textit{Office of Government and} \\
\textit{Community Relations}\\
\textit{Lawrence Berkeley National Laboratory}\\
Berkeley, CA, USA \\
esanchez@lbl.gov}
}

\maketitle

\begin{abstract}
The 2024 Quantum Computing, Math and Physics Camp (QCaMP) for Students was a 4-week long summer camp aimed at introducing high school students to quantum concepts and careers, including applications spanning quantum computing, sensing, and communication.  The program ran for 7 hours/day, Monday--Friday, July 1--26, and included hands-on modules and activities, professional development, and project-based learning.  Here we provide details on the camp curriculum and outcomes based on pre and post knowledge and attitudes assessments.
\end{abstract}


\section{Introduction}
There have been significant investments in quantum applications in recent years, including \$38.6b worldwide in federal funding alone~\cite{Qureca2023}. Additionally, the Quantum Economic Development Consortium (QED-C) continues to report strong job numbers~\cite{QEDC2025}.  Today's quantum jobs are multidisciplinary, drawing talent from a broad range of STEM fields and at a variety of degree levels, with less than half of jobs requiring a PhD~\cite{PhDs}. The Quantum Information Science and Technology (QIST) Workforce Development National Strategic Plan points out that exposure to QIST at the high school or undergraduate level is limited, likely leaving talented students to study other fields due to a lack of awareness~\cite{QISTWDNSP2018}.  The 2024 QCaMP built upon a prior 1-week long camp~\cite{Ivory2023} to provide a unique 4-week opportunity for high school students to learn QIST concepts, applications, and hands-on skills from researchers through engaging modules, professional development, and project-based learning, with the goal of providing students with awareness of and pathways into careers in QIST\@.  This expansion was made possible due to funding from the Department of Energy, Office of Science, Workforce Development for Teachers and Scientists, Pathway Summer Schools initiative.

We describe new elements of the QCaMP curriculum in Section~\ref{sec:Curriculum}, evaluate the program's outcomes in Section~\ref{sec:Outcomes}, and outline planned improvements for future iterations of QCaMP in Section~\ref{sec:Outlook}.

\section{Curriculum Overview}
\label{sec:Curriculum}

Similar to previous iterations of QCaMP, QIST researchers from national laboratories collaboratively developed the curriculum around topics shown in Table~\ref{Curr}. The modules were developed to be accessible to a broad range of students by avoiding any prerequisite math, physics, or computer science courses. Modules also incorporated experiential and active, hands-on learning wherever possible. In addition, quick multiple choice and short answer questions were incorporated throughout the modules to check for student understanding, allowing instructors to revisit any concepts as needed. With the expansion of the camp to 4 weeks, students had sufficient time to participate in project-based learning (Section~\ref{Projects}, and professional development and lab tours (Section~\ref{sec:ProfessionalDevelopment}.  These additional camp components gave students the opportunity to gain  hands-on skills, science communication skills, and a deeper understanding of QIST research topics.  

\begin{table*}[htbp]
\caption{QCaMP for Students 2024 Curriculum}
\begin{center}
\begin{tabular}{|c||c|c|c|c|c|}
\hline
& \textbf{Monday} & \textbf{Tuesday} & \textbf{Wednesday} & \textbf{Thursday} & \textbf{Friday}\\
\hline\hline
\multirow{2}{*}{\textbf{Wk1-AM}} & \multirow{2}{*}{Clifton Strengths} & \multirow{2}{*}{Classical Bits and Gates} & \multirow{2}{*}{IBM Quantum Composer} & 4th of July & 5th of July \\
& & & & OFF & OFF \\
\hline
\multirow{2}{*}{\textbf{Wk1-PM}} & Icebreakers and & Gates as Matrices & Hadamard and & 4th of July & 5th of July \\
& Pre-Assessments & (Section~\ref{Lesson:GatesasMatrices}) & Superposition & OFF & OFF\\
\hline\hline
\multirow{2}{*}{\textbf{Wk2-AM}} & Polarization/Malus's Law & Bloch Sphere & Entanglement, & Superconductivity & Linear Algebra \\
& (Section~\ref{Lesson:MalussLaw}) & (Section~\ref{Lesson:BlochSphere}) & \$ vs. Tiger & (Section~\ref{lesson:SuperconductingQubits}) & (Section~\ref{Lesson:PythonPrimerandLinearAlgebra})\\
\hline
\multirow{2}{*}{\textbf{Wk2-PM}} & \multirow{2}{*}{Growth Mindset} & \multirow{2}{*}{Probability} & Python Primer & Measurement and QKD & \multirow{2}{*}{Peer-to-Peer Teaching} \\
&  &  & (Section~\ref{Lesson:PythonPrimerandLinearAlgebra}) & (Section~\ref{Lesson:Measurement} \&~\ref{Lesson:QKD}) & \\
\hline\hline
\multirow{2}{*}{\textbf{Wk3-AM}} & Quantum Sensing & Noise and Error Correction & \multirow{2}{*}{Wave/Particle Duality} & Quantum Annealing & \multirow{2}{*}{Career Panel} \\
& (Section~\ref{Lesson:QuantumSensing}) & (Section~\ref{Lesson:NoiseandErrorCorrection}) & & (Section~\ref{Lesson:QuantumAnnealing}) & \\
\hline
\multirow{2}{*}{\textbf{Wk3-PM}} & \multirow{2}{*}{Lab Tours} & \multirow{2}{*}{Projects} & \multirow{2}{*}{Projects} & \multirow{2}{*}{Projects} & \multirow{2}{*}{Projects} \\
&  &  &  &  & \\
\hline\hline
\multirow{2}{*}{\textbf{Wk4-AM}} & \multirow{2}{*}{Projects} & \multirow{2}{*}{Projects} & \multirow{2}{*}{Projects} & \multirow{2}{*}{Post-Assessments} & Symposium and \\
&  &  &  &  & Lab Tours\\
\hline
\multirow{2}{*}{\textbf{Wk4-PM}} & \multirow{2}{*}{Projects} & \multirow{2}{*}{Projects} & \multirow{2}{*}{Presentation Prep} & \multirow{2}{*}{Presentation Prep} & Poster Session \\
&  &  &  &  & and Awards\\
\hline
\end{tabular}
\label{Curr}
\end{center}
\end{table*}

\subsection{Lessons}
\label{sec:Lessons}

Here we describe in more detail the newly developed or modified lessons in the 2024 camp.  Note that some modules were kept from previous iterations of QCaMP with little modification: Classical Bits and Gates, IBM Quantum Composer, Wave/Particle Duality, Polarization, Hadamard and Superposition, Probability and Entanglement and \$~vs.\ Tiger~\cite{Economou2020}. Descriptions of these modules can be found in~\cite{Ivory2023}.

\subsubsection{Gates as Matrices}
\label{Lesson:GatesasMatrices}
Due to the additional 3 weeks in 2024's camp compared to prior camps, we had more opportunity to tie key quantum concepts to underlying mathematics.  Following the introduction of bits, binary, gates, and circuits, we introduced linear algebra concepts including vectors (which represent the state of a bit) and matrices (which represent gates).  Students were taught the basics of multiplying matrices and vectors in order to determine the outcome of a circuit.  Students practiced this math by hand during the initial lesson, and revisited it during the later Linear Algebra and Python lessons.

\subsubsection{Malus's Law}
\label{Lesson:MalussLaw}
The Polarization module has been part of QCaMP since 2022 and has been taught with a conceptual approach to accommodate students who have not yet taken trigonometry. With additional time added due to the 4-week expansion, there was an opportunity to include Malus's Law as a follow-up activity to the Polarization module and introduce students to quantitative aspects of polarization that they observed. The Virtual Quantum Optics Laboratory (VQOL)~\cite{LaCour2022} and Desmos~\cite{Desmos2025} were utilized to create an accessible online experiment where students were tasked with verifying Malus's Law. The students were first introduced to sine, cosine, and tangent functions, before the instructor shared Malus's Law and explained each component of it. Next, students set up a simple apparatus on VQOL with a laser, neutral density filter, polarizer, and a detector. Students were tasked with recording the result from the detector each time they changed the angle of the polarizer. A Desmos Graphing Calculator template was shared with students so they could graph in real-time their results from VQOL and compare them to Malus's Law. Finally, students discussed possible sources of error for their experiment.

\subsubsection{Bloch Sphere}
\label{Lesson:BlochSphere}
We introduced the Bloch sphere as a visual and conceptual tool to deepen understanding of quantum states. We began by revisiting the classical bit, which takes on discrete values of 0 or 1. To help ground this concept visually, we compared the classical bit to a position on a stick with two ends. We then shifted to a richer analogy: a ship navigating on a globe. Its location and direction is determined by three continuous coordinates—longitude, latitude, and heading—mirroring the more complex structure of a quantum state.

Students were each given small globes and worked through hands-on exercises identifying locations using coordinates. This tactile approach helped anchor the abstract ideas in physical experience. We then labeled specific points on the globe and introduced their quantum analogs: the North Pole as $|0\rangle$, the South Pole as $|1\rangle$, and the points where the prime meridian crosses the equator representing $|+\rangle$ and $|-\rangle$.

With this foundation, we formally introduced the Bloch sphere as the quantum version of the globe. We explained how the Euler angles (longitude, latitude, and heading) correspond to numerical parameters of a quantum state. Students revisited the state $|\psi\rangle = \alpha|0\rangle + \beta|1\rangle$ (introduced previously in Hadamard and Superposition lesson~\cite{Ivory2023} and learned how this abstract expression maps to a specific position on the Bloch sphere. To complete the session, we explored how quantum state evolution can be viewed as rotation on the Bloch sphere. This visual metaphor—rotating the globe around various axes—provided a concrete way to think about how quantum gates transform states. This exercise with the globe helped demystify what it means to manipulate a qubit.

\subsubsection{Python Primer and Linear Algebra}
\label{Lesson:PythonPrimerandLinearAlgebra}
The Python Primer introduced students to classical programming fundamentals while highlighting key use cases of computing within science and engineering: automation, simulation, and visualization. Throughout, the lesson reinforced concepts from earlier modules through interactive examples and exercises of increasing complexity. After learning and practicing necessary syntax, students automated an exercise from the Classical Bits and Gates module by implementing a function to convert text into binary. Then, students performed numerical experiments related to the Probability module by flipping a virtual coin and plotting statistical errors under an increasing number of coin flips. Finally, students used the Qiskit library~\cite{qiskit2024} to simulate single-qubit quantum gates and visualize their operation on the Bloch sphere, reviewing concepts from the Hadamard and Superposition and Bloch Sphere modules.

The Linear Algebra lesson aimed to bridge the gap between fundamental mathematical concepts and their application in quantum mechanics and quantum computing. The module's primary goal was to equip students with a working knowledge of vectors, linear transformations, and matrix representations, specifically within the context of two-dimensional real vectors. Employing a blended approach of theoretical explanations, pen-and-paper exercises, and hands-on Python coding, the curriculum progressed from basic vector definitions and operations to simulating a ``flat qubit'' (i.e. linear transformations from the special orthogonal group SO(2)). By focusing on intuitive geometric interpretations and practical computational exercises, students gained a concrete understanding of how linear algebra provides the mathematical language for describing and manipulating quantum states, ultimately laying the groundwork for more advanced quantum computing concepts.

The Python and Linear Algebra modules both used Python notebooks hosted on Google Colab to avoid time-consuming software installation and to enable an integrated lesson format where written explanation, example code, and programming exercises all appear in one place.

\subsubsection{Superconducting Qubits}
\label{lesson:SuperconductingQubits}
This module provided a phenomenological introduction to the fundamental concepts of superconductivity, with an eye towards how superconducting circuits could be used for quantum information science. We began by relating the various phases of electrical conductivity to phases of matter that students may be more familiar with. This approach allowed for a direct analogy between superfluidity as a phase of matter and superconductivity as a phase of electrical conductivity, with both phenomena existing at extremely cold temperatures. To give students a perspective on the temperatures required for our experiments, we depicted a temperature scale with different units, with specific notable examples of materials and transitions at various temperatures on the scale. After this contextualization, we delved into the properties of superconductors, giving a historical perspective on the seminal discoveries, including perfect conduction and the Meissner effect. We walked through the evolving approaches to theoretically understanding the microscopic explanation of superconductivity over time, including the London, Gizburg-Landau, and BCS formulations. During this historical perspective, we described the various elements and compounds that exhibit superconductivity, along with their critical temperatures.

Following the fundamental and historical overview, we explored applications of superconductors, including lossless power transmission, levitating trains, and powerful electromagnets. This discussion led to our focus application: superconducting quantum information processors. We primed the students by explaining how the superconducting circuit approach is quite similar to how conventional computers are built. We then described the role of fundamental circuit elements (such as resistors, inductors, and capacitors) in storing and dissipating electromagnetic energy, leading into the introduction of resonator circuits, which can store electromagnetic energy in oscillating electric and magnetic fields. These circuits are the key elements of superconducting circuits, along with the critical additional element:  the Josephson junction. We finished the presentation with some microscope images of various superconducting devices.

After the presentation, the students participated in a lab to explore the Meissner effect. The goal was to realize a levitating superconducting train car on a magnetic train track. The high-$T_C$ superconducting disk was cooled with liquid nitrogen, allowing the students to experience a cryogenic experiment. As the liquid nitrogen evaporated, the superconductor warmed, and the students witnessed the phase transition out of the superconducting state as the train stopped levitating. 

\subsubsection{Measurement}
\label{Lesson:Measurement}
We began our exploration of quantum measurement by reviewing the difference between discrete and continuous observables. To illustrate this distinction clearly, we used familiar examples like counting eggs in a basket versus weighing a car, making the abstract concepts tangible for participants. Then, we introduced the idea of a measurement procedure by identifying its key elements: the object being measured, the measuring device, the observer, and the overall procedure.

We emphasized a crucial difference between classical and quantum measurement. Classical measurements were highlighted as passive processes that do not affect the object being measured. In contrast, quantum measurement actively influences the object, fundamentally changing its state. Revisiting the earlier introduced concept of quantum states, participants were reminded of the wavefunction expressed as $|\psi\rangle = \alpha|0\rangle + \beta|1\rangle$. Quantum measurement was then defined as the collapse of this quantum state into one of two possible outcomes, humorously described as the ``Universe choosing a path''. To help students grasp this complex idea intuitively, we drew analogies to everyday life decisions. Each decision we make shapes our future, but quantum mechanics intriguingly suggests alternative paths might simultaneously exist, leading naturally to discussions of the multiverse concept. This analogy proved particularly effective, sparking thoughtful discussions among students.

Ultimately, our key message was clear: unlike classical measurement, quantum measurement is inherently interactive, altering the state of the object being measured. This subtle yet profound difference underscored the unique and often counterintuitive nature of quantum phenomena.

\subsubsection{Quantum Key Distribution}
\label{Lesson:QKD}
The Quantum Key Distribution (QKD) module aimed to introduce a new application of quantum technology,  illustrate this application's ramifications for security, and show how this application may impact the students' day-to-day lives. This lesson began with a basic history of cryptography including the examples of Caesar Cipher, smoke signals, and the enigma machine used in WWII\@. We then introduced classical cryptography protocols, including RSA~\cite{Rivest1978} and Advanced Encryption Standard (AES)~\cite{NIST2001}. Next, we showed the differences in speed of deciphering the RSA protocol using a classical computer compared to a quantum computer~\cite{Shor1994}. We showed the quantum advantages and highlighted reasons why QKD is an important research area for our security and a potential future research area for the students.

The lesson culminated in teaching a QKD protocol called BB84~\cite{Bennett&Brassard1984} and showing how using quantum concepts (specifically superposition, polarization, and bases) to generate an encryption key will guarantee security. Students' understanding of this topic was further developed with a hands-on activity using coin boxes~\cite{TuftelLab2024} that helped mimic a quantum process of sending and receiving quantum information and then measuring the quantum information to generate a guaranteed secure key. This activity served as an analogy of the BB84 protocol.

\subsubsection{Quantum Sensing}
\label{Lesson:QuantumSensing}
In addition to modules on quantum computing and quantum communication, we also introduced quantum sensing as a QIST application.  This module introduced students to defect energy levels in nitrogen vacancy (NV) centers in diamond and how their spin-dependent photoluminescence can be used to sense magnetic fields.  The lesson started by introducing these concepts and applications and ended with a demonstration of the NV-based photoluminescence microscope that students would have the opportunity to build in the associated Quantum Sensing project (See Section~\ref{Proj:QuantumSensing}).

\subsubsection{Noise and Error Correction}
\label{Lesson:NoiseandErrorCorrection}
The Noise and Error Correction module introduced students to the fundamental concept of noise as the source of computational errors and explored error correction techniques in both classical and quantum computers. We began with a historical overview of how noise disrupted computers during computation, which led to the emergence of error correction in classical computers. The simplest approach to error correction, the repetition code, was introduced to students through interactive activities, where they create bit strings, disrupt the bit strings (thus generating errors), and correct the errors as a team to gain an intuitive understanding of how redundancy can protect information. Concepts in modern error correction techniques such as code distance and Hamming code were then presented, and a worksheet strengthened the students' understanding of these new concepts.

Building upon classical error correction concepts, the module transitioned to quantum error correction, where not only bit-flip errors, but also phase-flip errors occur. Students first saw how classical error correction principles extend to quantum computing through the quantum repetition code, followed by the similarity between the Hamming code and the surface code in quantum computers (parity check). Finally, students learned about state-of-the-art implementations of the surface code in leading quantum computing platforms.


\subsubsection{Quantum Annealing}
\label{Lesson:QuantumAnnealing}
While based on the same underlying principles as the circuit model, quantum annealing~\cite{Kadowaki1998:quant-anneal} presents a fundamentally different view of quantum computing. The goal of this new module was to broaden students' perspective of what constitutes quantum computing. The module began by highlighting similarities to what the students have already learned—qubits, superposition, entanglement, etc.—but then explaining that quantum annealing is a more specialized technology: it is used to solve combinatorial optimization problems.

As combinatorial optimization is likely to be an unfamiliar subject to high-school students, the module first presented some examples such as the traveling salesperson routing problem; scheduling of machines, people, tasks; social network analysis of Romeo and Juliet; and decomposition of graphs (lines and edges) into equal parts or communities. It then introduced a hands-on exercise we called the ``sit/stand game''.  Students counted off ``A'', ``B'', ``C'', ``A'', ``B'', ``C'', \ldots\ to form teams of three. They were then presented with the scoring system (see Table~\ref{SitStand}) and given five minutes to decide who on their team should sit or stand in order to score the most points. Students learned that  (a) it is possible to have multiple equally optimal solutions (in this case, stand–sit–stand and stand–stand–sit) and (b) while this game's $2^3 = 8$ possibilities can be exhaustively enumerated in a few minutes, a similar game for teams of 2000 people would require an incomprehensibly large search space of $2^{2000}$ possibilities. Yet a quantum annealer could find a good, but not necessarily optimal, solution to this larger problem in a split second.

\begin{table}[htbp]
\caption{SIT-STAND: Scoring in the sit-stand game}
\begin{center}
\begin{tabular}{|c|c|}
\hline
\textbf{Position} & \textbf{Score}\\
\hline
A is standing & Lose 2 points\\
\hline
B is standing & Gain 0 points\\
\hline
C is standing & Lose 1 point\\
\hline
A and B are both standing & Gain 3 points\\
\hline
A and C are both standing & Gain 4 points\\
\hline
B and C are both standing & Lose 4 points\\
\hline
\end{tabular}
\label{SitStand}
\end{center}
\end{table}

The students were then introduced to quantum annealer hardware concepts such as topology, qubits, couplers, and connectivity. They watched two videos from D\nobreakdash-Wave on ``How the quantum annealing process works''~\cite{DwaveVid1} and ``Physics of quantum annealing: Hamiltonian and eigenspectrum''~\cite{DwaveVid2}, each followed by a few interactive questions to check for understanding.

The next part of the module walked students through problem formulation for a quantum annealer: problems must be expressed as quadratic unconstrained binary optimization (QUBO) problems~\cite{Glover2019:qubo}. This is a difficult topic, so a large amount of time was devoted to it, including more interactive questions to test the students' knowledge. At the end, students were shown how the sit/stand game maps directly to a QUBO and were also introduced to the max-cut problem. In addition to these concepts, students were shown Python code for implementing and running a quantum-annealing max-cut solver using (classical) simulated annealing.

Slides for formulating a map four-coloring problem (the map of Australia) were provided for reference. Future versions of this module will include these slides in the module proper.

\subsection{Projects}
\label{sec:Projects}

Students worked in groups on projects during the last two weeks of QCaMP\@. Student groups received guidance from their mentors, including daily check-ins, but the projects were largely driven by students, allowing them to develop skills in collaboration, experimental design, troubleshooting, and synthesizing their results to be communicated clearly through a scientific-style poster. Table~\ref{Projects} introduces the projects, the main skills/concepts involved, and the source of the project equipment if commercially available kits were used.
Each of the projects is described in the following subsections.

\begin{table*}[htbp]
\caption{Projects for Project-Based Learning in Weeks 3--4}
\begin{center}
\begin{tabular}{|c|l|}
\hline
\textbf{Project} & \textbf{Skills/Concepts}\\
\hline
3D Printed QKD~\cite{3DQKD} & Skills: experimental design, Arduino programming, optical setup \\
Section~\ref{Proj:QKD}& Concepts: BB84 protocol, polarization, QKD\\
\hline
Thorlabs QKD~\cite{ThorlabsQKD} & Skills: optical setup and alignment, experimental design \\
Section~\ref{Proj:QKD} & Concepts: BB84 Protocol, polarization, QKD\\
\hline
TeachSpin Two Slits~\cite{TeachSpin} & Skills: experimental design, data collection, analysis, and visualization \\
Section~\ref{Proj:DoubleSlit}& Concepts: interference patterns, Wave-particle duality, Young's double-slit experiment\\
\hline
Thorlabs Quantum Eraser~\cite{ThorlabsQE} & Skills: optical setup and alignment, experimental design \\
Section~\ref{Proj:QuantumEraser}& Concepts: wave-particle duality, quantum superposition\\
\hline
Planck's Constant~\cite{Carolina} & Skills: data collection and analysis, error estimation \\
Section~\ref{Proj:PlancksConstant}& Concepts: wave-particle duality, quantization of energy\\
\hline
Quantum Sensing & Skills: optical setup and alignment, experimental design, data collection, analysis\\
Section~\ref{Proj:QuantumSensing}& Concepts: spin, quantization of energy, magnetic-field sensing\\
\hline

Quantum Simulation & Skills: Python programming, IBM Q Composer, data analysis and visualization\\
Section~\ref{Proj:QuantumSimulation}&Concepts: quantum simulation, variational quantum eigensolver\\
\hline
Quantum Machine Learning & Skills: Python programming, machine learning\\
Section~\ref{Proj:QuantumMachineLearning}& Concepts: quantum simulation, variational quantum eigensolver \\
\hline
Quantum Games & Skills: Java programming, project management, game design\\
Section~\ref{Proj:QuantumGames}&Concepts: bits \& gates\\
\hline
Quantum Annealing & Skills: Python programming, converting visual puzzles into boolean code\\
Section~\ref{Proj:QuantumAnnealing}&Concepts: quadratic unconstrained binary optimization problems, quantum simulation\\
\hline
\end{tabular}
\label{Projects}
\end{center}
\end{table*}

\subsubsection{Quantum Key Distribution}
\label{Proj:QKD}
Two groups of students applied what they learned from the QKD lesson to a tabletop experiment about the BB84 protocol. One group used used the ThorLabs kit~\cite{ThorlabsQKD} and involved learning how to set up and align lasers, half wave plates, beam splitters, and detectors. The other group used the University of Waterloo kit~\cite{3DQKD}, which involved programming an arduino and aligning lasers with half wave plates and detectors. Through these experiments, students furthered their grasp of the BB84 protocol and how it generates a guaranteed secure key. The projects pushed the students to understand how QKD could impact information security as well as the quantum phenomenology driving this experiment. Students came to QCaMP with various levels of prior experience (many minimal to none) with quantum physics; however, by the end of the project, all students in the groups were able to explain the BB84 protocol in multiple ways.

\subsubsection{Double-Slit Experiment}
\label{Proj:DoubleSlit}
This project builds upon a prior lesson introducing students to wave interference and wave particle duality~\cite{Ivory2023}.  Using a the ``Two-Slit Interference, One Photon at a Time'' kit from TeachSpin~\cite{TeachSpin}, students were able to explore the interference pattern resulting from classical light passing through a single and double slit in a quantitative way by mapping out the light intensity as a function of position.  Furthermore, this kit enables the reduction of light source intensity such that only one photon at a time passes through the single or double slit apparatus.  A photomultiplier tube and photon counting unit allows students to quantify the single photon results.  Students were challenged to determine mathematically if they were truly in the single photon regime.

\subsubsection{Quantum Eraser}
\label{Proj:QuantumEraser}
The quantum eraser project and the Planck's Constant project, which is described in Section~\ref{Proj:PlancksConstant}, were paired together to provide a hands-on approach for students to deepen their understanding of wave-particle duality and quantum measurement. The quantum eraser project used the Thorlabs Quantum Eraser Demonstration Kit~\cite{ThorlabsQE}, which enabled the students to conduct a variation of Thomas Young's double-slit experiment. Students constructed and aligned a Mach-Zehnder interferometer on an optical breadboard and incorporated polarizers to mark and erase optical path information. This experiment was done with a visible continuous wave laser and therefore provided a quantum analogy where the hypotheses could be defined and experimental results could be explained in terms of both classical and quantum physics.

\subsubsection{Planck's Constant}
\label{Proj:PlancksConstant}
In this research project, which was paired with the quantum eraser project (see Section~\ref{Proj:QuantumEraser}), students carried out a set of experiments with four light emitting diodes (LEDs) to ultimately determine an experimentally extracted value for Planck's constant~\cite{Carolina}. They set up a circuit to power the LEDs, measured the threshold voltages, and used diffraction gratings to determine the LED wavelengths. From this information, the students made plots of the photon energy as a function of frequency to extract Planck's constant. Sources of error were discussed, such as the LEDs emitting only approximately a single wavelength. The objective of this project was to help students gain a deeper understanding of the particle nature of light and energy quantization while encouraging them to conduct hands-on experiments.

\subsubsection{Quantum Sensing}
\label{Proj:QuantumSensing}
In this project, students built a photoluminescence microscope, used it to measure the magnetic fields generated by a permanent magnet as a function of distance from a nitrogen-vacancy (NV) rich diamond, extracted the magnetization density of the neodymium magnet, and finally compared their result to known values of neodymium magnetization density. This first-of-its-kind quantum sensing project relied on having components for a photoluminescence microscope, an easily implemented quantum sensing controller (QICK-DAWG~\cite{qick-dawg}), and instructor support. Students were provided with a microscope assembly guide and Jupyter notebooks to run measurements and fit data. The students learned optical system assembly/alignment, Python fitting, and how to optically detect magnetic resonance of NVs to extract the applied magnetic field. Students came away from the module understanding how externally applied perturbations (magnetic field) can affect discrete spin energy levels of a quantum system (the nitrogen-vacancy).

\subsubsection{Quantum Simulation}
\label{Proj:QuantumSimulation}
 A Python-based quantum simulation project introduced students to prospective quantum computing applications within chemistry and materials science by simulating a hydrogen dimer with the variational quantum eigensolver (VQE). Starting from matrix representations of single-qubit and two-qubit gates, students implemented a Python emulation of the VQE circuit~\cite{OMalley2016} and optimized the variational parameter to obtain the ground-state energy. Repeating this procedure over a range of possible interatomic separation distances, they determined the equilibrium bond length of the molecule. Finally, students used IBM Q to execute some of the VQE circuits on available quantum hardware and examined the effects of noise on their predictions.

\subsubsection{Quantum Machine Learning}
\label{Proj:QuantumMachineLearning}
The goal of this project was similar to the above Quantum Simulation project (see Section~\ref{Proj:QuantumSimulation}), where students used  VQE to solve for the ground-state energy of a hydrogen dimer~\cite{OMalley2016}.  In this project, students compared different optimization algorithms for updating the variational parameter---including gradient descent, R-Adam, and Adam optimizers---in order to determine which algorithm identifies the ground-state energy in the fewest number of iterations.  Students used Google Colab, including Pennylane and supporting libraries for the Python code.

\subsubsection{Quantum Games}
\label{Proj:QuantumGames}
The goal of this project was to develop a video game to help make quantum computing more approachable. The focus of the game was solving logic puzzles involving gates. Similar to Tetris, bits fall towards quantum gates and the player must determine the correct output based on the given inputs in a limited time. The puzzles get more challenging as the game progresses. The game was made without any starter code and had a working prototype by the end of QCaMP.

\subsubsection{Quantum Annealing}
\label{Proj:QuantumAnnealing}
\begin{figure}[htbp]
\centerline{\includegraphics[scale=.6]{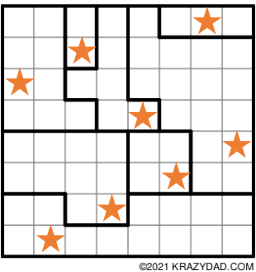}}
\caption{An example of a solved Star Battle puzzle from krazydad.com~\cite{NYTTwoNotTouch} distributing 8 stars such that there is one in every column, row, and region but without any two stars touching horizontally, vertically, or diagonally.}
\label{starbattle}
\end{figure}

Building upon the Quantum Annealing module (see Section~\ref{Lesson:QuantumAnnealing}), the goal of this project was for students to get hands-on experience programming a quantum annealer to solve a sudoku-like logic puzzle called Star Battle (carried by the New York Times as Two Not Touch). A Star Battle puzzle consists of an $N{\times}N$ grid of cells partitioned into irregularly shaped regions. The player---or in our project's case, the quantum annealer---is asked to fill $N$ cells with stars such that (a) every row, every column, and every irregular region contains exactly one star and (b) no two stars touch horizontally, vertically, or diagonally.  See Figure~\ref{starbattle} for an example of a solved Star Battle puzzle.

For the project, students were asked to use D\nobreakdash-Wave's Ocean library to program a Python solution to a given Star Battle puzzle. They needed to determine how to represent the grid as a collection of Boolean variables, how to express ``exactly 1 of k variables must be True'' as a QUBO, how to express ``at most 1 of k variables must be True'' as a QUBO, and how to implement each of these components for a quantum-annealing solution.

Star Battle relies on the same QUBO primitives as map coloring, giving us an opportunity to address those ideas from the Quantum Annealing lesson (see Section~\ref{Lesson:QuantumAnnealing}). The project was fundamentally programming-heavy and required more Python training than what the students received in the Python Primer described in Section~\ref{Lesson:PythonPrimerandLinearAlgebra}. Unfortunately, D\nobreakdash-Wave prohibits providing accounts on their cloud platform, Leap, to minors. Without access to D\nobreakdash-Wave hardware, the students tested their project on a local simulator based on (classical) simulated annealing. This approach worked, but it was not as satisfying as running on a real quantum annealer. Despite all these challenges, the students managed to complete the project and produce and present a poster describing their solution approach. 

\subsection{Professional Development and Lab Tours}
\label{sec:ProfessionalDevelopment}

In addition to the modules and projects described in Sections~\ref{sec:Lessons} and~\ref{sec:Projects}, professional development opportunities were added throughout the program to support students' college and career readiness. Topics included Clifton strengths assessment, building a resume and LinkedIn profile workshop, growth mindset, applying and paying for college, as well as career panels representing not just professionals from QIST fields but also from other STEM fields. In addition, in preparation for the poster symposium at the end of the program, students participated in sessions where they learned how to create an effective scientific poster.

Students also had opportunities to visit different labs based on their location. After the Superconducting Qubits lesson (see Section~\ref{lesson:SuperconductingQubits}), students were given a tour of the Advanced Quantum Testbed (AQT) at Berkeley Lab, either in-person for the students at Berkeley or through a guided virtual platform using the AQT's publicly accessible website. Students at Berkeley also toured two user facilities at Berkeley Lab, the National Energy Research Scientific Computing Center (NERSC) and the Advanced Light Source (ALS). Students in the New Mexico cohorts participated in a virtual tour of the QSCOUT lab~\cite{QSCOUT2025,QSCOUT2025tour}, as well as the VECSEL lab and the Quantum Optics and Quantum Information lab at the University of New Mexico.

\subsection{Ties to Standards}

Modules were linked with Quantum Information Science (QIS) Key Concepts for K--12 Framework from the National Q-12 Education Partnership~\cite{Q12}. Table~\ref{Standards} lists the standards covered in the High School Physics, Computer Science, and Mathematics frameworks for each module. Note that the IBM Q Composer module is not included as the purpose of this module was to introduce the IBM Q Composer as a tool and revisit practice problems covered during Bits \& Gates.

\begin{table}[htbp]
\caption{Lessons and linked QIS Key Concepts}
\begin{center}
\begin{tabular}{|c|c|}
\hline
\textbf{Lesson} & \textbf{QIS Key Concepts} \\
\hline
Bits \& Gates & HS.Phys.4.1, HS.CS.4b.1 \\
\hline
Hadamard \& Superposition & HS.Phys.7.1, HS.Math.4a.1 \\
\hline
\multirow{2}{*}{Polarization} & HS.Phys.2.4, HS.Phys.3.2, HS.Phys.3.3,\\
& HS.Phys.3.4, HS.Phys.4.2, HS.Math.2a.1 \\
\hline
Bloch Sphere & HS.Phys.4.1, HS.Phys.4.2AP1, HS.Phys.4.3 \\
\hline
Probability and Statistics & HS.Math.2b1.2 \\
\hline
\multirow{2}{*}{Entanglement, \$ vs.\ Tiger} & HS.Phys.5.1, HS.Phys.5.2, HS.Phys.5.3,\\
& HS.CS.5a.1, HS.CS.7b.2 \\
\hline
Superconducting Qubits & HS.Phys.7.7, HS.Phys.7.8 \\
\hline
Measurement and Bases & HS.Phys.3.1, HS.Phys.3.2, HS.Math.3b.1 \\
\hline
\multirow{2}{*}{Quantum Key Distribution} & HS.Phys.8.1a-g, HS.CS.8b.1,\\
& HS.CS.8b.2, HS.CS.8b.3 \\
\hline
\multirow{2}{*}{Python / Linear Algebra} & HS.Phys.4.2AP1, HS.CS.2a.1, HS.CS.2a.3,\\
& HS.CS.2a.4, HS.Math.2b2.1 \\
\hline
Quantum Sensing & HS.Phys.9.1, HS.Phys.9.2 \\
\hline
\multirow{2}{*}{Noise \& Error Correction} & HS.Phys.6.1, HS.Phys.6.2, \\
& HS.Phys.6.5, HS.CS.2c.1 \\
\hline
\end{tabular}
\label{Standards}
\end{center}
\end{table}

\section{Outcomes}
\label{sec:Outcomes}

To evaluate the effectiveness of the 2024 QCaMP, we administered pre and post knowledge and attitudes assessments to our student participants.  The knowledge assessment gauged student ability to correctly answer concept questions related to each of the lesson topics and showed an average score of 19\% correct prior to the camp and 84\% correct at the completion of the camp.  The attitudes assessment gauged student sentiments such as perceived understanding of subjects, sense of belonging, interest, and other reflections on the camp execution.

Additional details about the methodology (Section~\ref{sec:outcomes:methods}), student participants (Section~\ref{sec:outcomes:participants}) and outcomes from the knowledge (Section~\ref{sec:outcomes:knowledge}) and attitudes (Section~\ref{sec:outcomes:attitudes}) assessments are found below.

\subsection{Methods}
\label{sec:outcomes:methods}

The evaluation team worked with the camp leadership to design a survey that addressed key outcomes the camp experiences were designed to impact, including attitudes about quantum science, career interests, and teamwork. The pre-survey was administered at the start of the camp, and the post-survey was administered at the end of the camp. The post-survey included questions measured on Likert scales and also had opportunities for students to provide comments to open-ended questions about what they enjoyed, what can improve, and other facets of their experience. The survey was built electronically, and data were collected using Research Electronic Data Capture (REDCap).~\cite{Harris2009}

Data collected through closed-ended questions were analyzed using frequencies and descriptive statistics. Data collected from open-ended questions were analyzed using thematic, inductive coding~\cite{Creswell2016} whereby the evaluators read through participants' responses and generated common themes that emerged.

\subsection{Participant Information}
\label{sec:outcomes:participants}

The 2024 cohort consisted of 24 students in Berkeley, CA, 14 students in Albuquerque, NM, and 3 students in Santa Fe, NM\@. This year's cohort was required to be 16 years old by the first day of the camp.  The only prerequisite was Algebra 1, so the camp was made accessible to students without any prior advanced math, physics, chemistry, or even programming classes.  
Table~\ref{Grades} shows the distribution of students by grade for the 2024 cohorts, as indicated by student assessment responders (N=40).
Additionally, 50\% of NM students and 44\% of CA students attended Title I schools, and 78\% of NM students and 84\% of CA students attended public schools.

\begin{table}[htbp]
\caption{Participant Grade Levels from Assessment Responses}
\begin{center}
\begin{tabular}{|c|c|c|c|c|c|}
\hline
\textbf{9th } & \textbf{10th} & \textbf{11th } & \textbf{12th } & \textbf{College} & \multirow{2}{*}{\textbf{Total}}\\
\textbf{Grade} & \textbf{Grade} & \textbf{Grade} & \textbf{Grade} & \textbf{Freshmen} & \\
\hline
1 & 1 & 19 & 14 & 5 & 40\\
\hline
\end{tabular}
\label{Grades}
\end{center}
\end{table}

\subsection{Knowledge-Based Assessment}
\label{sec:outcomes:knowledge}

\begin{figure}[htbp]
\centerline{\includegraphics[scale=.8]{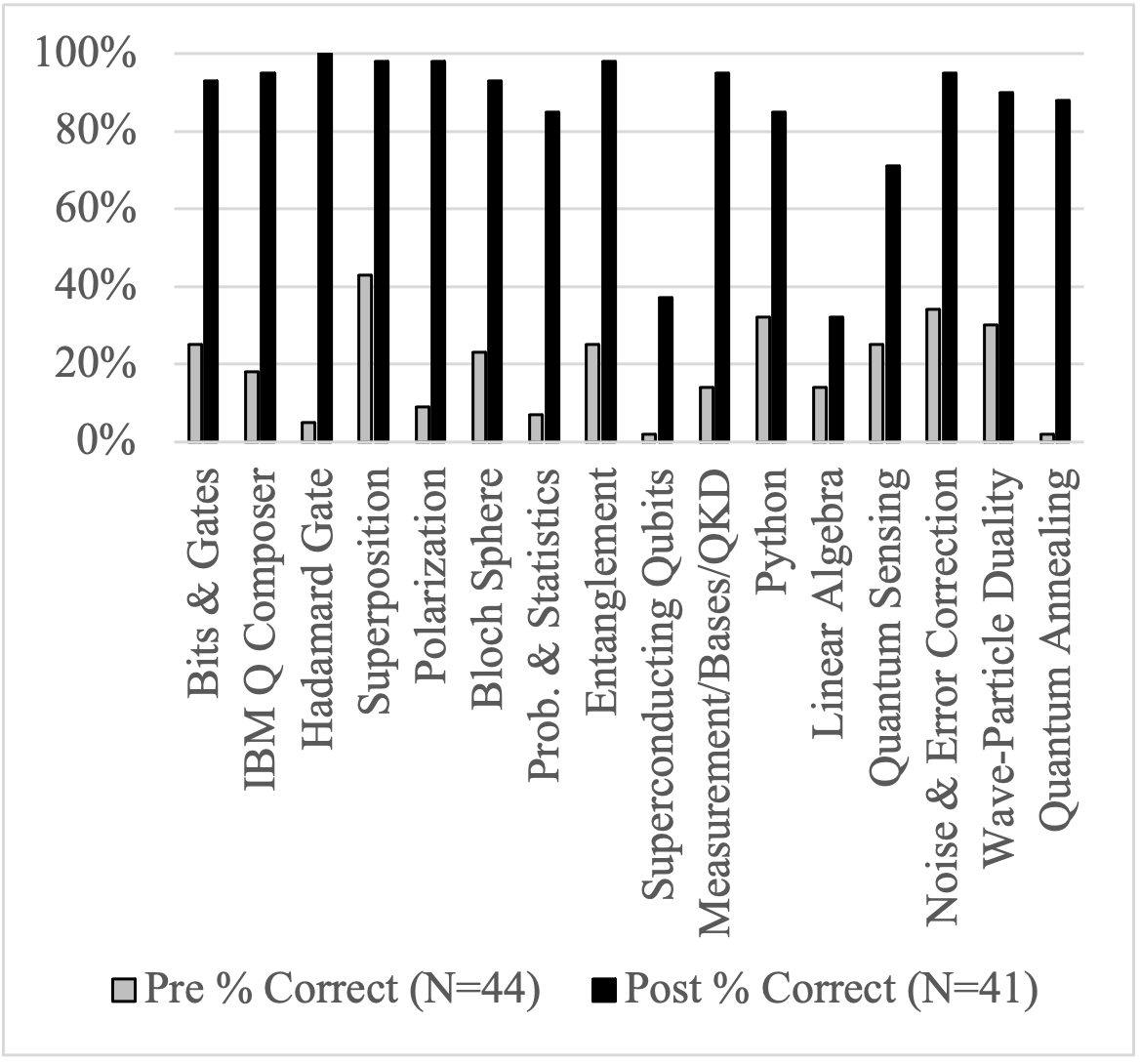}}
\caption{A knowledge-based assessment was administered to students before and after the camp.  Each lesson contributed a single multiple choice question to the assessment.}
\label{knowledge}
\end{figure}

In order to identify areas of strength and improvement needed for future iterations of the curriculum, students completed a pre-program and post-program knowledge assessment. The assessments contained multiple choice questions with students having the option to indicate ``I am not familiar with this concept'', which was not counted as correct. This option is included to prevent correct responses that result from guessing. The results from the assessments are shown in \mbox{Figure~\ref{knowledge}}. Overall, there is an increase in the percentage of correct answers across all lessons, with all but 3 lessons reaching over 80\% correct. Two categories that did not reach over 50\% correct, Superconducting Qubits and Linear Algebra, had a knowledge-based question that was not adequately covered within the lesson.  These lessons, as well as the Quantum Sensing module, will be updated for future iterations of QCaMP to address student misconceptions. 

\subsection{Attitudes Assessment}
\label{sec:outcomes:attitudes}

In addition to the knowledge-based assessment, an attitudes assessment was completed before and after the camp to gauge students' self-reported understanding of topics like quantum science, math, computing, and physics; interest in STEM careers; sense of belonging, skill development, personal recommendation, overall value of the camp; etc.  Table~\ref{Attitudes} shows a selection of the questions posed to students on a 5-point Likert scale ranging from poor to excellent (questions 1--6), none to exceptional (questions 7--14), or strongly disagree to strongly agree (questions 14--20, only asked in the post-assessment).  The Pre and Post columns give average responses to each question.  Pre and Post results were analyzed via a T-Test for two independent means with a two-tailed hypothesis to identify three statistically significant results, as indicated by the bolded questions and calculated P values.

\begin{table*}[htbp]
\caption{Attitudes Assessment Change from Pre to Post}
\begin{center}
\begin{tabular}{|l|c|c|c|c|}
\hline
\textbf{Question} & \textbf{Pre} & \textbf{Post} & \textbf{Change} & \textbf{P-Value}\\
\hline
\textbf{1.  My understanding about quantum science is} & 1.725 & 3.2 & +1.475 & \textbf{$<$0.0001}\\
\hline
\textbf{2.  My understanding about computing is} & 2.7 & 3.3 & +0.6 & \textbf{$<$0.05}\\
\hline
3.  My understanding about mathematics is & 3.3 & 3.6 & +0.3 &  $>$0.05 \\
\hline
\textbf{4.  My understanding about physics is} & 2.5 & 3.05 & +0.55 & \textbf{$<$0.05}\\
\hline
5.  My understanding that failing and mistakes are a part of the scientific process is & 4.3 & 4.375 & +0.075 & $>$0.05 \\
\hline
6.  My understanding of how I contribute to teams that I am a part of is & 3.875 & 3.95 & +0.075 & $>$0.05 \\
\hline
7.  My comfort in asking questions about material is & 3.55 & 3.6 & +0.05 & $>$0.05 \\
\hline
8.  My comfort in working on a team with my peers is & 3.825 & 3.9 & +0.075 & $>$0.05 \\
\hline
9.  My interest in pursuing a STEM career is & 4.35 & 4.45 & +0.1 & $>$0.05 \\
\hline
10. My interest in pursuing a quantum-related career is & 3.15 & 3.1 & -0.05 & $>$0.05 \\
\hline
11. My sense of belonging to people with similar interests to me is& 3.743 & 3.7 & -0.043 & $>$0.05 \\
\hline
12. My sense of belonging in the field of science is & 3.925 & 3.8 & -0.125 & $>$0.05 \\
\hline
13. My comfort with making mistakes on tasks is & 3.45 & 3.85 & +0.4 & $>$0.05 \\
\hline
14. My interest in taking a class in quantum computing, math, and/or physics this school year is & 4.205 & 4.25 & +0.045 & $>$0.05 \\
\hline
15. When I get back to school this year, I am going to tell all my friends about QCaMP. & n/a & 4.1 & n/a & n/a \\
\hline
16. I am interested in participating in additional QCaMP programs. & n/a & 4.075 & n/a & n/a \\
\hline
17. I feel like I can succeed in my STEM classes. & n/a & 4.3 & n/a & n/a\\
\hline
18. I have support from a mentor/teacher/other in the science community to help me succeed in a STEM career. & n/a & 4.2 & n/a & n/a\\
\hline
19. I feel like I belong in the field of science. & n/a & 4.128 & n/a & n/a\\
\hline
20. Overall, QCaMP was a valuable experience. & n/a & 4.65 & n/a & n/a\\
\hline
\end{tabular}
\label{Attitudes}
\end{center}
\end{table*}

In questions that do not indicate a statistically significant change from pre to post, it is still insightful to consider the values.  For example, despite the lack of prerequisites for this camp, students indicated a strong interest in pursuing a STEM career and taking additional classes in quantum computing, math, and/or physics in the upcoming school year. In the post assessment, students also indicated strong interest in additional QCaMP programs, having strong mentorship support, feelings of belonging in science, ability to succeed in STEM classes, and that QCaMP was a valuable experience.

Students were also given the opportunity to answer open-ended questions related to various aspects of the camp.  Here, they commented on activities they most enjoyed, as well as ways to improve the camp.  Some student responses are captured below:

\begin{itemize}
    \item ``Have the lessons be more hands on so that they're not as overwhelming and they get a chance to interact with the material they're learning.''
    \item ``I think expanding the research portion of the camp, as well as expanding the amount of concepts taught would be very valuable.''
    \item ``I enjoyed the environment and getting to know the people from QCaMP the most, and being familiar with the people there made me more comfortable.''
\end{itemize}

\section{Outlook}
\label{sec:Outlook}

Expanding QCaMP from a 1-week long summer program to a 4-week long summer program allowed us an opportunity to further engage with students, including a broader number of topics, which required lesson development from scratch in many cases, and project-based learning. For future iterations of QCaMP, curriculum and schedule changes will be implemented based on student feedback. More hands-on labs and activities will be integrated with each lesson module, including more opportunities for students to work collaboratively throughout the camp. To increase the scope of the content covered, an additional module introducing quantum chemistry and atomic qubits will be incorporated. There will also be an expanded time period for students to work on the group research projects.

As students develop an interest in quantum information science and technology through QCaMP, we aim to share future opportunities that will support them in gaining more experiences and skills in this field. QIST internships and programs are shared with students continuously after QCaMP. 
Despite efforts by both Berkeley Lab and Sandia National Laboratories to develop continuous pathways into QIST, there remain gaps in opportunities at the high school, community college, and undergraduate levels~\cite{2025NQEC}.  Additional programs, including summer camps, workshops, technician training programs, etc.\ can help fill these gaps and ensure a pipeline of quantum-ready talent.

\section*{Acknowledgment}

We acknowledge leadership activities performed by Jake Douglass, Bethany Cannon, and Deb Menke at Sandia National Laboratories, Faith Dukes of the K--12 Programs at Lawrence Berkeley National Laboratory, Eileen Dawson at the Quantum Systems Accelerator, and Jeff Nelson at the Center for Integrated Nanotechnology (CINT); instruction of previously developed lessons by Ermal Rrapaj and Mohan Sarovar; logistical assistance provided by Milad Marvian, Ivan Deutch, and Dwight Zier at the University of New Mexico; camp facilitation by Emma South in Santa Fe, NM and Bailee Rusconi, Nicolas Canastuj-Velasco, and Jaime Choy in Berkeley, CA; curriculum assistance by Mojgan Haghanikar; camp facilitation in Albuquerque, NM and project development and oversight by Joan Arrow; and UNM lab tours provided by Hermann Kahle and Elohim Becerra.


QCaMP is funded under the Department of Energy Office of Science Workforce Development for Teachers and Scientists Pathway Summer Schools initiative (``QCaMP: Building a Quantum-Ready Workforce''). 
Additional support is acknowledged from Sandia National Laboratories and Lawrence Berkeley National Lab.  Some of the work presented in this article was supported by the National Security Education Center (NSEC) Informational Science and Technology Institute (ISTI) under the Laboratory Directed Research and Development (LDRD) program of Los Alamos National Laboratory project number 20240479CR-IST\@.  Los Alamos National Laboratory is operated by Triad National Security, LLC for the National Nuclear Security Administration of the U.S. Department of Energy (contract no.~89233218CNA000001).  Some of this research used resources of the National Energy Research Scientific Computing Center (NERSC), a Department of Energy Office of Science User Facility funded by the US DOE (Contract No.~DE-AC02-05CH11231).  Some of this work was performed, in part, at the Center for Integrated Nanotechnologies, an Office of Science User Facility operated for the U.S. Department of Energy (DOE) Office of Science.


Sandia National Laboratories is a multi-mission laboratory managed and operated by National Technology \& Engineering Solutions of Sandia, LLC (NTESS), a wholly owned subsidiary of Honeywell International Inc., for the U.S. Department of Energy's National Nuclear Security Administration (DOE/NNSA) under contract DE-NA0003525. This written work is authored by an employee of NTESS\@. The employee, not NTESS, owns the right, title and interest in and to the written work and is responsible for its contents. Any subjective views or opinions that might be expressed in the written work do not necessarily represent the views of the U.S. Government. The publisher acknowledges that the U.S. Government retains a non-exclusive, paid-up, irrevocable, world-wide license to publish or reproduce the published form of this written work or allow others to do so, for U.S. Government purposes. The DOE will provide public access to results of federally sponsored research in accordance with the DOE Public Access Plan.

\renewcommand{\doitext}{doi:~}  

\vspace{12pt}

\end{document}